\documentclass[a4paper]{jpconf}
\usepackage{graphicx}
\begin{document}
\title{Link between laboratory and astrophysical radiative shocks}

\author{C Michaut$^1$, E Falize$^{1,2}$, C Cavet$^1$, S Bouquet$^{1,2}$, M Koenig$^3$, T Vinci$^{2,3}$, B Loupias$^3$}

\address{$^1$ LUTH, Observatoire de Paris, CNRS, Universit\'e Paris Diderot ; 5 Place Jules Janssen, 92190 Meudon, France}
\address{$^2$ D\'epartement de Physique Th\'eorique et Appliqu\'ee, CEA/DIF, 91680 Bruy\`eres-le-Ch\^atel, France}
\address{$^3$ LULI, Ecole Polytechnique, CNRS, CEA, UPMC, 91128 Palaiseau cedex, France}

\ead{claire.michaut@obspm.fr}
%%%%%%%%%%%%%%%%  
\begin{abstract}
This work provides analytical solutions describing the post-shock structure of radiative shocks growing in astrophysics and in laboratory. The equations including a cooling function $\Lambda \propto \rho^{\epsilon} P^{\zeta} x^{\theta}$ are solved for any values of the exponents $\epsilon$, $\zeta$ and $\theta$. This modeling is appropriate to astrophysics as the observed radiative shocks arise in optically thin media. In contrast, in laboratory, radiative shocks performed using high-power lasers present a radiative precursor because the plasma is more or less optically thick. We study the post-shock region in the laboratory case and compare with astrophysical shock structure. In addition, we attempt to use the same equations to describe the radiative precursor, but the cooling function is slightly modified. In future experiments we will probe the PSR using X-ray diagnostics. These new experimental results will allow to validate our astrophysical numerical codes.
\end{abstract}
%%%%%%%%%%%%%%%%  
\section{Introduction}
In this work both radiative shocks (RS) arising in astrophysics and those generated in laboratory are studied. As these RS are involved in all stages of stellar evolution (accretion shocks, pulsating stars, supernovae and interstellar medium), accurate modeling is needed and, therefore, experimental results are used to validate our codes. Until now our team performed RS ex\-periments~\cite{Michaut:ASS:07,Koenig:POP:06,Bouquet:PRL:04} using the high-power laser at Laboratoire d'Utilisation des Lasers Intenses (\'Ecole Poly\-technique). In laboratory, the plasma can be considered at local thermodynamical equilibrium and optically thick or intermediate~\cite{Drake:Book:06}. With specific radiative hydrodynamics codes~\cite{Michaut:ASS:07}, we study these RS structures since at high-Mach numbers ($M$) they exhibit precursor.\\
On the other hand in astrophysics, the plasma is often optically thin~\cite{Bertschinger:APJ:86} and radiation escapes without interaction with the surrounding material. Its role can be modeled~\cite{Bertschinger:APJ:86} by a cooling function $\Lambda(\rho,P)$. The main objective of analytical modeling suggested here is to predict the extension of the optically thin cooling zone behind the RS front. We consider the equations presented in \cite{Chevalier:APJ:82}, but in this paper they are solved analytically for any $\Lambda$ proportional to a power law of $\rho$ and $P$. In this case only the post-shock region (PSR) is structured by cooling like for Polars where a RS arises for magnetic white dwarf accreting neighboring star material.\\
In addition, we attempt to calculate the precursor length of steady laboratory RS. Based on Drake's work~\cite{Drake:Book:06}, we switch the cooling function by a equivalent system which represents the radiation flux propagating towards the precursor. 
In next experiments scheduled in '08 on LIL (Bordeaux, France), we plan to probe PSR using X-ray diagnostics. With analytical solutions of the above astrophysical model, we can predict the structure of PSR for these experiments. In the same way, experimental results of the post-shock cooling will allow to validate our astrophysical codes, since by confrontation with analytical results we can determine the main physical process {\it i.e.} the value of the exponents in $\Lambda$.\\
%%%%%%%%%%%%%%%%  
\section{Theoretical modeling}
\begin{figure}[h]
\begin{minipage}{14pc}
\includegraphics[width=14pc]{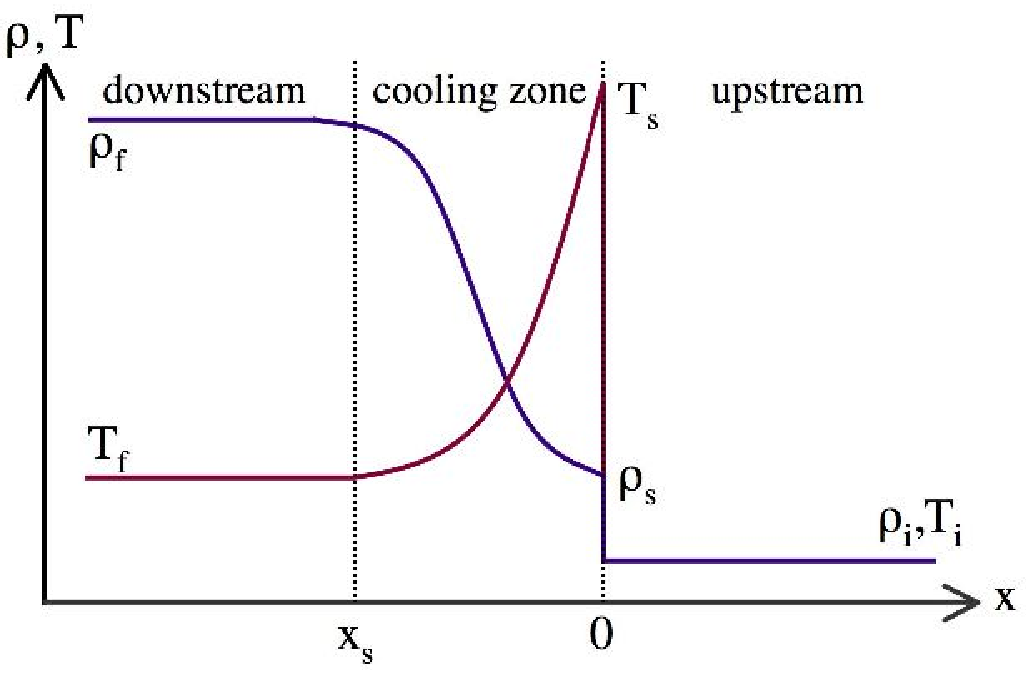}
\caption{\label{scheme1} Schematic structure of a RS in an optically thin medium.}
\end{minipage}\hspace{2pc}%
\begin{minipage}{14pc}
\includegraphics[width=14pc]{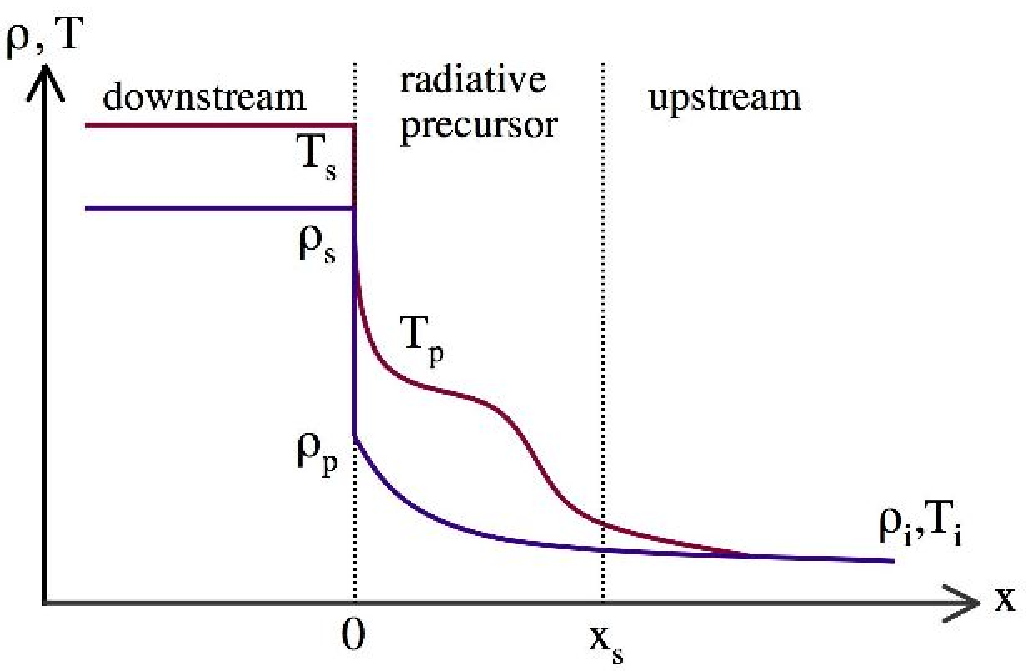}
\caption{\label{scheme2} Schematic structure of a weakly RS in an optically thick medium.}
\end{minipage} 
\end{figure}

We are interested to describe the structure of the cooling PSR (see Fig.~\ref{scheme1}) and to predict its extension $x_{s}$.
We consider a one dimensional (1D) stationary RS where propagation is along $x$ and $\rho$, $v$, $T$ represent respectively mass density, velocity, temperature and $P$ is the pressure. With a cooling function $\Lambda$, the Euler equations are ($\gamma$ is the polytropic index of ideal gas):
\begin{equation} \label{Eqsys}
   \frac{d}{dx} \left[ \rho v \right] = 0 \;\; ;  \;\; 
   \frac{d}{dx} \left[ \rho v^{2} + P \right] = 0  \;\;  ;  \;\; 
    v \left[ \frac{dP}{dx} - \gamma \frac{P}{\rho} \frac{d \rho}{dx}  \right] = - (\gamma - 1) \,  \Lambda (\rho,P) \;\; .
\end{equation}  
Since system~(\ref{Eqsys}) contains $\Lambda$, the calculation of the gradient zone is possible. According to previous work~\cite{Drake:Book:06,Chevalier:APJ:82}, the cooling function may have the following form ($\Lambda_{0}$ is constant):
\begin{equation} \label{Eq1cool}
   \Lambda (\rho,P)  =  \Lambda_{0} \rho^{\epsilon} P^{\zeta} {(x+x_{0})}^{\theta} \; ,
\end{equation} 
where $\epsilon$, $\zeta$ and $\theta$ are three exponents, the value of which is defined according to the considered physical processes. In order to solve the system, we have to introduce in Eq.~(\ref{Eq1cool}) a minimal value $x_{0}$ to prevent a singularity at $x=0$ and we have to define the boundary conditions.\\
%%%%%%%%%%%%%%%%  
\subsection{Boundary conditions}
First of all, we introduce a new dimensionless variable $\xi = x/x_{s}$ where $x_{s}$ represents the cooling PSR length. Then the inverse of the local compression ratio is $\eta(\xi)=\rho_{i} / \rho(\xi)$. As the integration is done from $\eta(0)$ to $\eta(\xi)=0$, the solution diverges with a value of $\rho$ going to infinity. Therefore the boundary conditions are given by the value of $\eta$ at the shock front ($\xi=0$) and they depend on the type of considered shock at the other end ($\xi=1$).  We have to cut the curve at $x=x_s$ as soon as $\rho$, $v$ or $T$ satisfies chosen boundary conditions. At the shock front side, the compression ratio for a stationary shock is given by Rankine-Hugoniot conditions versus $M$: 
\begin{equation} \label{EqRH}
  \frac{\rho_{s}}{\rho_{i} } = \frac{(\gamma + 1) M^{2}}{(\gamma - 1) M^{2} + 2} = \frac{1}{\eta(0)}\;
\end{equation} 
As an example, the rear shock will be modeled by wall conditions in the case where the RS represents the plasma falling onto a white dwarf in Polar type cataclysmic variable with $T=0$~\cite{Chevalier:APJ:82} (see Polar RS in section~\ref{sec:polar}). In the case of a bow shock ahead a young stellar jet, boundary conditions are pressure equilibrium~\cite{Blondin:APJ:89}, {\it i.e.} the rear pressure must be equal to the Mach disk pressure given by observations. Further discussions about boundary conditions are in \cite{Ramachandran:MNRAS:06}.\\
We also extend this analytical solutions to predict the evolution of the precursor in the case of a strong RS in optically thick medium. 
The radiative flux is replaced with the function $\Lambda (\rho,P)$ in standard equations and boundary conditions are given by the gas density in the precursor. Now let continue with the resolution of these equations in PSR, in optically thin medium.\\
\subsection{Analytical solutions}
In this section, we can not detail each calculation step which will be the subject of a next article. %But we note we have also solved the whole sytem for a cooling function which modelize two different physical processes. 
%First of all, we introduce a new space variable without dimension $\xi = x/x_{s}$ in which $x_{s}$ represents the post-shock length.
Considering the appropriate boundary conditions (see Eq.~(\ref{EqRH})), the pressure becomes: 
\begin{equation}
   P(\xi) = \rho_{i} v_{s}^{2} (1+ \frac{1}{\gamma M^{2}} - \eta(\xi)) \,\; {\rm and} \,\;   v(\xi) = v_{s} \times \eta(\xi) \; .
 \end{equation}    
 where $v_{s}$ is the shock velocity. Under these variable changes, the first two equations of (\ref{Eqsys}) are fulfilled and give the same energy conservation than Bertschinger~\cite{Bertschinger:APJ:86}. These solutions, written under an implicit form, are: 
%%%%
 \begin{eqnarray} \label{Eqsol}
 \nonumber  (\gamma+1) \eta^{(\epsilon + 2)}   & 
                       \times {\left[ 1 + \frac{1}{\gamma M^{2}} \right]}^{- \zeta}   &
                       \times \frac{\Gamma(\epsilon + 2)}{\Gamma(\epsilon + 3)} 
                       \times \,  _{2}F_{1} \Big( \zeta, \epsilon+2;  \epsilon+3; \eta / \left[ 1 + \frac{1}{\gamma M^{2}} \right] \Big)  \\
 \nonumber  -  (\gamma) \eta^{(\epsilon + 1)}   &
                       \times \left[ 1 + \frac{1}{\gamma M^{2}} \right]^{1 - \zeta}   &
                       \times \frac{\Gamma(\epsilon + 1)}{\Gamma(\epsilon + 2)} 
                       \times \,  _{2}F_{1} \Big( \zeta, \epsilon+1;  \epsilon+2; \eta / \left[ 1 + \frac{1}{\gamma M^{2}} \right] \Big) \\
                      + A_{0} & = &  \Bigg\{ \begin{array}{ll}
                                                            \kappa_{0} \ln(\xi+x_{0}/x_{s}) &  {\rm if} \;  \theta = -1 \\ 
                                                            \kappa_{0}/{\theta+1} (\xi+x_{0}/x_{s})^{\theta+1} &  {\rm if} \;  \theta \neq -1 \end{array}  \; \; .
\end{eqnarray}
Up to now, this problem was solved for only five cases~\cite{Laming:PRE:04} and Eq.~(\ref{Eqsol}) provides the general analytical solution (including the former five) for system~(\ref{Eqsys}).
\section{Astrophysical application: Polars or AM Her objects} \label{sec:polar}
Polars stars are a class of cataclysmic variables (binary system with an accreting compact object) in which the strong magnetic field of the white dwarf (first star) completely dominates the accretion flow of the system. Indeed the matter of the second star follows the magnetic lines of the white dwarf up to its poles and falls down on its surface in an accretion RS. In PSR, a cooling layer forms that is described  with a cooling function as seen before. To produce this cooling zone, the involved physical process is an optically thin Bremsstrahlung radiation~\cite{Fabian:MNRAS:76} with $\epsilon=3/2$, $\zeta=1/2$ and $\theta=0$ in Eq.~(\ref{Eq1cool}).  We obtain with our analytical solution the same expression for the PSR length as in numerical studies~\cite{Wu:APJ:94, Kylafis:APJSS:82}.
%%%%%%%
\section{Applications to laboratory radiative shock experiments}
\subsection{Precursor length evaluation}
With Eqs.~(\ref{Eqsys}) we calculate the precursor length for laboratory RS (see notations in Fig.~\ref{scheme2}). In this case $\Lambda$ represents the radiative flux escaping from the shock front towards the precursor, therefore the required parameters are evaluated as $\Lambda_{0}=627932$, $\epsilon=-1$, $\zeta=2$, since $\theta$ keeps free and here we find $\theta=1.41$ according to modified black-body radiation~\cite{Drake:Book:06}. As $\epsilon=-1$, the analytical development differs from Eq.~(\ref{Eqsol}). Under these approximations the precursor length $x_{s}$ is found equal to 200 $\mu m$ which is in agreement with the steady-state limit in previous experiments and simulations~\cite{Michaut:ASS:07}. This result is interesting because it strongly links an astrophysical analytical aspect to laboratory experiments.
\subsection{Future radiative shock experiment scheduled on LIL}
\begin{figure}[h]
\includegraphics[width=14pc]{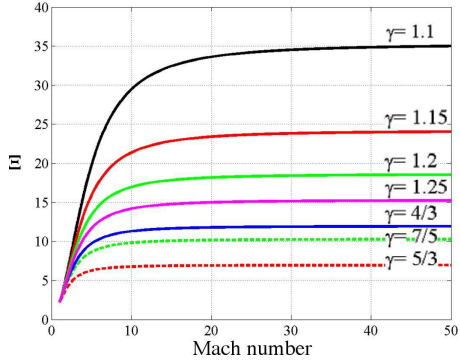}\hspace{2pc}%
\begin{minipage}[b]{14pc}\caption{\label{column} Accreted column density $\Xi=\int_{0}^{x_{s}} \rho(x)\, dx/\int_{0}^{x_{s}} \rho_i\, dx$, versus the Mach number for ($\epsilon$,$\zeta$)=(3/2,1/2) which corresponds to a cooling Bremsstralhung and for different value of $\gamma$.}
\end{minipage}
\end{figure}

We have solved and generalized calculations of the cooling PSR for any couple $(\epsilon,\zeta)$. To apprehend future experiments, another parameter which is extremely interesting to estimate is the accreted column density $\Xi$ (see Fig.~\ref{column}).
As $\Xi$ is directly related to the compression ratio and as we plan to measure the shock front density using X-ray diagnostic, we will be able to deduce the nature of the main physical process responsible for the cooling behind the front shock by comparing analytical and experimental results. This point is important to introduce the real physics in numerical simulation codes. The purpose of this calculation is to predict the cooling zone length behind the stationary RS. Since we expect to probe this region, we need to know future target size and diagnostic position.
%%%%%%%
\section{Conclusion}
We have generalized the analytical solutions of the model for any cooling function and compared them with previous numerical work~\cite{Wu:APJ:94} and with some trivial analytical solutions~\cite{Chevalier:APJ:82,Laming:PRE:04,Imamura:APJ:85}. Although we have recovered already known results, we have derived additional classes of solutions. This work provides directly pieces of information relevant to search for astrophysical objects or phenomena suitable for comparisons with (rescaled) data obtained experimentally. Based on recent Drake's work~\cite{Drake:Book:06}, we can apply same calculation to evaluate the length of the radiative precursor of laboratory radiative shocks. Moreover with equations of the post-shock zone we can prepare future experiments on lasers even more powerful than LIL as LMJ or NIF, under condition to reach a stationary shock regime. As a result, we can emphasize the occurring physical processes through the exponent determination in the cooling function. As a conclusion, we strengthen the connection between experimental and numerical studies on radiative shocks by introducing our analytical predictions.
%%%%%%%

%%%%%%%%

\section*{References}
\begin{thebibliography}{20}
\bibitem{Michaut:ASS:07} Michaut C, Vinci T, Boireau L {\it et al} 2007 {\it Astrophys. and Space Science} {\bf 307} 159
\bibitem{Koenig:POP:06}  Koenig M, Vinci T, Benuzzi-Mounaix A {\it et al} 2006 {\it Phys. Plasmas} {\bf 13} 056504
\bibitem{Bouquet:PRL:04} Bouquet S, Stehl\'e C, Koenig M {\it et al} 2004 {\it Phys. Rev. Lett.} {\bf 92} 225001
\bibitem{Drake:Book:06}  Drake R P {\it High Energy Density Physics: Fundamentals, Inertial Fusion and Experimental Astrophysics} (Springer Verlag, Berlin, 2006)
\bibitem{Bertschinger:APJ:86} Bertschinger E 1986 {\it Astrophys. J.} {\bf 304} 154
\bibitem{Chevalier:APJ:82} Chevalier R and Imamura J N 1982 {\it Astrophys. J.} {\bf 261} 543
\bibitem{Blondin:APJ:89} Blondin J M and Cioffi  D F 1989 {\it Astrophys. J.} {\bf 345} 853
\bibitem{Ramachandran:MNRAS:06} Ramachandran B and Smith M D 2006 {\it Mon. Not. R. Astron. Soc. }{\bf 366} 586
\bibitem{Laming:PRE:04} Laming J M 2004 {\it Phys. Rev. E} {\bf 70} 057402
\bibitem{Fabian:MNRAS:76} Fabian  A C, Pringle J E and Rees M J 1976 {\it Mon. Noc. R. Astr. Soc.} {\bf175} 43
\bibitem{Wu:APJ:94} Wu K, Chanmugan G and Shaviv G 1994 {\it Astrophys. J.} {\bf426} 664
\bibitem{Kylafis:APJSS:82} Kylafis N D and Lamb D Q 1982 {\it Astrophys. J. Suppl. Ser.} {\bf48} 239
\bibitem{Imamura:APJ:85} Imamura J N 1985 {\it Astrophys.  J.} {\bf 296} 128
\end{thebibliography}
\end{document}